\documentclass[twocolumn,prb,groupedaddress,amsmath,amssymb,floatfix]{revtex4}
\usepackage{graphicx}
\usepackage{dcolumn}
\usepackage{bm}
\begin{document}
\title{Coupling of Rotational Motion with Shape Fluctuations 
of Core-shell Microgels Having Tunable Softness}

\author{S. Bolisetty$^a$, M. Hoffmann$^a$, S. Lekkala$^a$, Th. Hellweg$^a$,  M. Ballauff$^a$, and L. Harnau$^b$}
\affiliation{$^a$Physikalische Chemie I, University of Bayreuth, 
D-95440 Bayreuth, Germany\\
$^b$Max-Planck-Institut f\"ur Metallforschung,  
Heisenbergstr.\ 3, D-70569 Stuttgart, Germany, 
\\and Institut f\"ur Theoretische und Angewandte Physik, 
Universit\"at Stuttgart, Pfaffenwaldring 57, D-70569 Stuttgart, Germany}

\date{\today}

\begin{abstract}
The influence of shape fluctuations on deformable thermosensitive microgels in aqueous 
solution is investigated by dynamic light scattering (DLS) and depolarized dynamic light 
scattering (DDLS). The systems under study consist of a solid core of polystyrene and a 
thermosensitive shell of cross-linked poly(N-isopropylacrylamide) (PNIPA) without and with 
embedded palladium nanoparticles. PNIPA is soluble in water, but has a lower critical solution 
temperature at \mbox{32\,$^{\rm o}$C} (LCST). Below the LCST the PNIPA shell is swollen. Here we find that besides translational 
and rotational diffusion, the particles exhibit additional dynamics resulting from shape 
fluctuations. This leads to a pronounced apparent increase of the rotational diffusion coefficient. Above the transition temperature the shell collapses and provides a rather tight envelope of the core. In this state the dynamics 
of the shell is frozen and the core-shell particles behave like hard spheres. 
A simple physical model is presented to capture and explain the essentials of the 
coupling of rotational motion and shape fluctuations. 
\end{abstract}
\maketitle

\section{Introduction}

In recent years, a lot of research has been focused on the preparation and investigation of  ``smart'' microgels consisting of a thermosensitive network of poly(N-isopropylacrylamide) (PNIPA).\cite{hu:98,juod:00,naya:05,dong:06,chu:07,lu:07a,cont:08} Major incentives for this research were possible applications in catalysis,\cite{lu:07a} photonics\cite{karg:07} or for the fabrication of responsive 
surface coatings.\cite{hell:08c} In particular, core-shell particles consisting of a polystyrene core onto which a 
thermosensitive network  is affixed present well-defined model colloids and exhibit 
polydispersities below $\pm$ 6\%.\cite{ding:98,dura:98,dura:98a,seel:01,hell:04}
Hence, besides their potential for applications these core-shell microgels 
are interesting model systems for studies of the flow behavior of concentrated colloidal suspensions.\cite{cra:06a,cra:08a} In this way core-shell microgels have become one of the best-studied class of polymer colloids.

In aqueous media, PNIPA exhibits a lower critical solution 
temperature (LCST) of about \mbox{32\,$^{\rm o}$C}. \cite{tana:85,shib:92,pelt:00,krat:01,wu:03}
Below this temperature the network is swollen by the solvent water whereas water is expelled 
from the microgel above the lower critical solution temperature. Hence, in the swollen state the
PNIPA shell of the core-shell microgels is expected to have a rather soft character. This implies the presence of network
breathing modes in this state. Up to now, there is evidence that the rheology of colloidal suspensions is strongly related to the dynamic properties of 
these often deformable objects. An important example in this context represents the flow 
of blood, containing deformable erythrocytes. Other important examples are liquid droplets,
emulsions, and vesicles. While experimental 
and theoretical studies have been devoted to the understanding of the dynamics of bending modes,\cite{huan:87,fara:90,hell:98,gang:94,kant:07,turi:08} the effect of shape fluctuations on the rotational diffusion coefficient of deformable objects has not been investigated yet 
despite the importance of the rotational degree of freedom for soft materials. 
\cite{abka:02,kant:05,nogu:07,lebe:07} Possible reasons may be sought in the lack of 
well-defined monodisperse model systems that can be studied by suitable experimental techniques.

\begin{figure}[t]
\begin{center}
\includegraphics[width=8cm, clip]{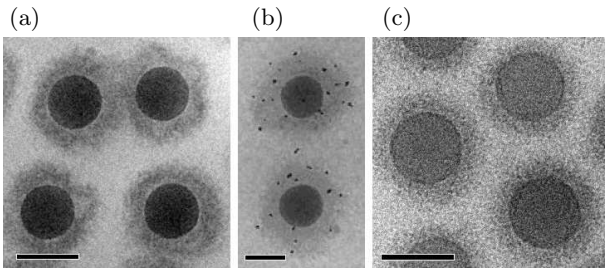}
\caption{CryoTEM micrographs of thermosensitive core-shell particles in aqueous solution.
The samples were maintained at \mbox{25\,$^{\rm o}$C} in (a), (b) and at 
\mbox{45\,$^{\rm o}$C} in (c) before vitrification. The dark core consists of polystyrene 
and the corona of PNIPA cross-linked with $N,N'$-methylenebisacrylamide. In (b)
palladium nanoparticles (black dots) are embedded in the PNIPA-shell.\cite{lu:06,lu:07b} The scale bars are 
\mbox{100 nm.}}
\label{fig1}
\end{center} 
\end{figure}

Here we study the translational and rotational motion of thermosensitive core-shell microgel
particles by depolarized dynamic light scattering (DDLS).\cite{bern:76,schm:90} The aim of the 
present work is a better understanding of the coupling of rotational motion and shape fluctuations. Such microgels are highly suitable for the present study because these particles have been extensively studied by scattering methods such as small-angle neutron scattering, small-angle X-ray scattering, and light scattering.\cite{crow:99,lu:07a,wu:03a,bern:03,stie:04,maso:05,bern:06a,bern:06b,zhao:08}
It has been demonstrated that metal nanoparticles can be embedded in the network 
of the shell. \cite{lu:06,lu:07a,lu:07b}   Moreover, recent work has shown that cryogenic transmission electron microscopy 
(Cryo-TEM) is well suited to study the structure and the shape of these particles 
in-situ:\cite{cra:06b,cra:08b} A thin film of the fluid dispersion containing the particles 
is shock-frozen and subsequently analyzed by transmission electron microscopy (TEM), no 
staining or any other preparatory step is necessary. Figure \ref{fig1} shows typical cryo-TEM
micrographs of dilute suspensions of such core-shell particles. The core consisting of polystyrene 
and the shell of cross-linked PNIPA is clearly visible. Figures \ref{fig1} (a) and (c) display 
the bare particles at different temperatures while Figure \ref{fig1} (b) shows a core-shell 
system where palladium nanoparticles are embedded in the shell. \cite{lu:06,lu:07b} The shape 
of the core-shell microgels shown in Figures \ref{fig1} (a) and (b) is  slightly asymmetric. 
This asymmetry must be traced back to thermal fluctuations. Therefore, we expect the dynamic 
properties of the core-shell system to be influenced by the shape fluctuations of the shell. \\
Depolarized dynamic light scattering (DDLS) is the method of choice for studying 
the problem at hand since this technique simultaneously probes the translational 
and rotational diffusion coefficient of optically anisotropic particles.\cite{bern:76,schm:90} DDLS has been applied to a number of anisometric particles in dilute solution.\cite{peco:68,degi:91,koen:00,eime:92} In general, spherical particles should not exhibit a signal in DDLS. However, the shape fluctuations of the core-shell microgels that are visible as frozen anisometry in the Cryo-TEM micrographs (see Fig. \ref{fig1}) should lead to an optical anisotropy of sufficient magnitude to give a finite depolarized signal in solution. In addition, even in the absence of these shape fluctuations network inhomogeneities due to an anisotropic distribution of crosslinking points will give rise to an additional contribution in the DDLS intensity autocorrelation function. This is observed indeed. On the other hand, the average shape of the particles is still spherical and rather simple models can still be applied. This conclusion can be derived from the fact that these particles 
have the phase diagram of hard spheres.\cite{cra:06a,cra:08a} \\

The paper is organized as follows: After the section Experimental we first give a brief survey on the theory of DDLS and 
develop a general scheme for the evaluation of data. In a second step we discuss 
the intensity autocorrelation functions for the microgels above and below their transition 
temperature in terms of a slow and a fast relaxation rate, the latter being related to the 
coupling of shape fluctuations and rotational motion. In the last section a simple statistical-mechanical model is presented that provides a qualitative explanation of the experimental data. 
%
%
%
%

\section{Experimental}
The synthesis and the characterization of the particles has been described previously. 
\cite{cra:06a,cra:06b,lu:07a} All solutions (0.05 wt \%) were prepared in 0.05 M KCl 
to reduce electrostatic particle interactions \cite{cra:06a,cra:08a} and filtered into 
dust free sample holders using 0.45 $\mu$m nylon filters. 

All experiments were carried out using the ALV/DLS/SLS-5000 compact goniometer system 
equipped with a He-Ne laser ($\lambda=632.8$ nm). The scattering cells (10 mm cylindric 
cuvettes, Hellma) were immersed in an index matching bath of cis-decaline which does not 
change the polarization plane of the laser light as e.g. toluene. For the DDLS experiment, 
the primary beam and the scattered light passed through a Glan-Thomson polarizer with an 
extinction coefficient better than 10$^{-5}$. The first polarizer guaranteed that mainly 
vertically polarized light impinges on the sample and the orientation of a second polarizer 
(analyzer) was carefully adjusted to a crossed position with the minimum scattered intensity. 
All radii of the particles $a$ given in the text were obtained by DLS.

\section{Intensity autocorrelation function}
The theory of dynamic light scattering has been presented in various treatises. 
\cite{bern:76,schm:90} Hence, we only review the equations necessary for 
this study. For an incident light wave traveling in the $x$ direction with a 
polarization vector in the $z$ direction the intensity of the scattered electric 
field can be written as 
\begin{eqnarray} 
I_s({\bf q},t)=I_{VV}({\bf q},t)+I_{VH}({\bf q},t)\,,
\end{eqnarray}
where the absolute value of the scattering vector ${\bf q}$ is given by 
$q=|{\bf q}|=(4\pi n/\lambda)\sin(\theta/2)$ in which $n$ is the refractive index 
of the medium. $\lambda$ is the incident wavelength and $\theta$ is the scattering 
angle. Pecora \cite{peco:65,peco:68} has given general expressions for 
$I_{VV}({\bf q},t)$ and $I_{VH}({\bf q},t)$ as 
\begin{eqnarray} 
I_{VV}({\bf q},t)\!\!&\sim&\!\!\int d{\bf r}\,d{\bf r}'\,
\left\langle \alpha_{zz}({\bf r}+{\bf r}',t) \alpha_{zz}({\bf r}',0)\right\rangle
e^{i{\bf q}\cdot{\bf r}}\,,
\\I_{VH}({\bf q},t)\!\!&\sim&\!\!\int d{\bf r}\,d{\bf r}'\,
\left\langle \alpha_{zy}({\bf r}+{\bf r}',t) \alpha_{zy}({\bf r}',0)\right\rangle
e^{i{\bf q}\cdot{\bf r}}\,, 
\end{eqnarray}
where $\alpha_{zz}({\bf r},t)$ and $\alpha_{zy}({\bf r},t)$ are the $zz$ 
and $zy$ elements of the fluid polarizability tensor. Experimentally accessible
quantities are the intensity autocorrelation functions $g^{(2)}_{VV}({\bf q},t)$
using dynamic light scattering (DLS) and $g^{(2)}_{VH}({\bf q},t)$ using DDLS. 
For photon counts obeying Gaussian statistics, the intensity autocorrelation functions 
are related to the electric field autocorrelation functions $g^{(1)}_{VV}({\bf q},t)$ 
and $g^{(1)}_{VH}({\bf q},t)$ according to
\begin{eqnarray}
g^{(2)}_{VV}({\bf q},t)&=&1+f_{VV} \left(g^{(1)}_{VV}({\bf q},t)\right)^2\,,
\\g^{(2)}_{VH}({\bf q},t)&=&1+f_{VH} \left(g^{(1)}_{VH}({\bf q},t)\right)^2\,, 
\end{eqnarray}
where $f_{VV}$ and $f_{VH}$ are dependent on the scattering geometry and are 
usually treated as adjustable parameters. The electric field correlation functions
can be calculated for various systems.

The core-shell particles in solution can change their position, orientation, and 
shape randomly by thermal agitation. For a dilute solution containing noninteracting
monodisperse spherical particles of radius $a$ the intensity autocorrelation functions 
are given by:
\begin{eqnarray} \label{eq1}
\sqrt{g^{(2)}_{VV}(q,t)-1}&=&
e^{-q^2D_T(a)t},
\\\sqrt{g^{(2)}_{VH}(q,t)-1}&=&
\frac{e^{-q^2D_T(a)t}\left(B(q,a)+e^{-6D_R(a,\alpha)t}\right)}
{B(q,a)+1}.\nonumber
\\&&
\label{eq2}
\end{eqnarray}
The translational and rotational diffusion coefficients read
\begin{eqnarray} \label{eq3}
D_T(a)&=&\frac{k_BT}{\eta}\frac{1}{6\pi a}\,,
\\D_R(a,\alpha)&=&\frac{k_BT}{\eta}\frac{1}{8\pi a^3}\alpha\,,  \label{eq4}
\end{eqnarray}
where the temperature $T$ and viscosity $\eta$ characterize the solvent and $\alpha=1$ 
for hard spheres. The parameter $B(q,a)$ in eq \ref{eq2} takes into account
possible contributions of polarized components to the intensity of the scattered light in 
the DDLS experiment due to a limited extinction ratio of the polarizer as discussed 
below. \\
As a new feature of the present evaluation, we have introduced the parameter $\alpha$ 
in eq \ref{eq4}. This parameter describes the difference of the real system from the hard 
sphere model, that is, $\alpha=1$: If $\alpha=1$, the rotational diffusion as well as the 
translational diffusion is fully described by a single parameter, namely the hydrodynamic 
radius $a$. If $\alpha \neq 1$, the rotational diffusion is coupled to an additional degree 
of freedom of the particle. Since $D_R(a,\alpha)$ scales with $a^{-3}$, possible deviations 
may hence be determined by precise DDLS-measurements.

\section{Results and Discussion}
\subsection{Microgels above the LCST}
We first discuss the microgel particles at temperatures above the volume transition. Here 
we expect the dynamics of the core-shell particles at \mbox{$T=45\,^{\rm o}$C} to 
be the same as those of hard spheres because the shell is fully collapsed under these 
conditions and provides a rather tight envelope 
of the core as is apparent from Figure \ref{fig1} (c). \cite{cra:06b}
The tight and nearly homogeneous shell visible above the transition temperature 
can be traced back to the fact that the particles are synthesized at high 
temperatures (80 $^\circ$C). \cite{ding:98} It must be kept in mind that the interaction 
between the particles becomes attractive above the transition temperature and the 
particles may coagulate slowly. \cite{cra:06a} However, the small concentrations used 
in the present DDLS experiments prevent this coagulation.

\begin{figure}[t!]
\begin{center}
\includegraphics[width=7.5cm,clip]{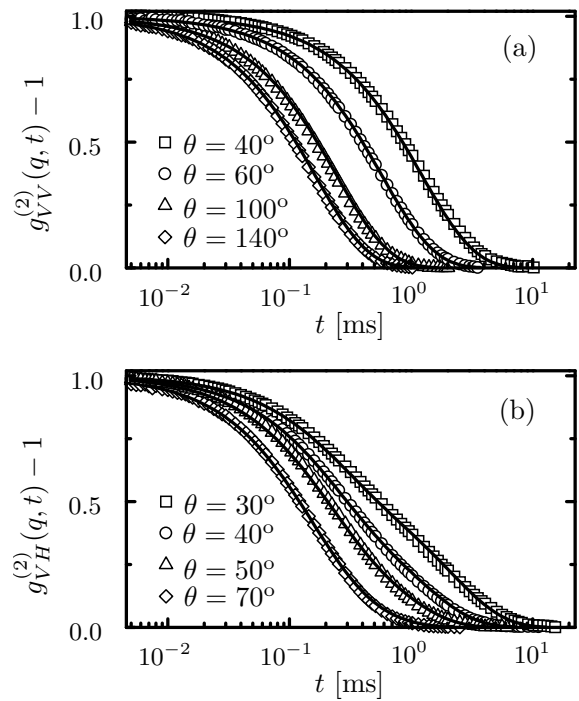}
\vspace*{-0.4cm}
\caption{(a) [(b)] DLS [DDLS] intensity autocorrelation functions $g^{(2)}_{VV}(q,t)$ 
[$g^{(2)}_{VH}(q,t)$] (symbols) of the core-shell particles containing palladium 
nanoparticles at \mbox{$T=45\,^{\rm o}$C} (see Figure \ref{fig1} (c)) together with the 
calculated results for monodisperse hard spheres (lines) according to 
eqs \ref{eq1} - \ref{eq4} with \mbox{$a=78$ nm} and \mbox{$\alpha=1$}.
In (a) and (b) the scattering angle $\theta$ increases from right to left.}
\label{fig2}
\end{center} 
\end{figure}

Figure \ref{fig2} displays examples of 
measured and calculated intensity autocorrelation functions of the core-shell particles 
containing palladium nanoparticles at \mbox{$T=45\,^{\rm o}$C}. In the calculations the 
model parameters \mbox{$\eta=0.601\times 10^{-3}$ Ns/m$^2$}, \mbox{$a=78$ nm}, and 
\mbox{$\alpha=1$} have been used. From both the translational and the rotational diffusion 
coefficient the same hydrodynamic radius of the microgel particles can be calculated (78 nm). 
The full agreement between the experimental data and the 
calculated results demonstrates that the theoretical approach according to
eqs \ref{eq1} - \ref{eq4} for hard spheres is indeed appropriate for the microgel 
particles at high temperature. Moreover, it 
indicates that the residual polydispersity does not disturb the measurements.
%
%
%
%

\subsection{Microgels below the transition}
\begin{figure}[t!]
\begin{center}
\includegraphics[width=7.5cm,clip]{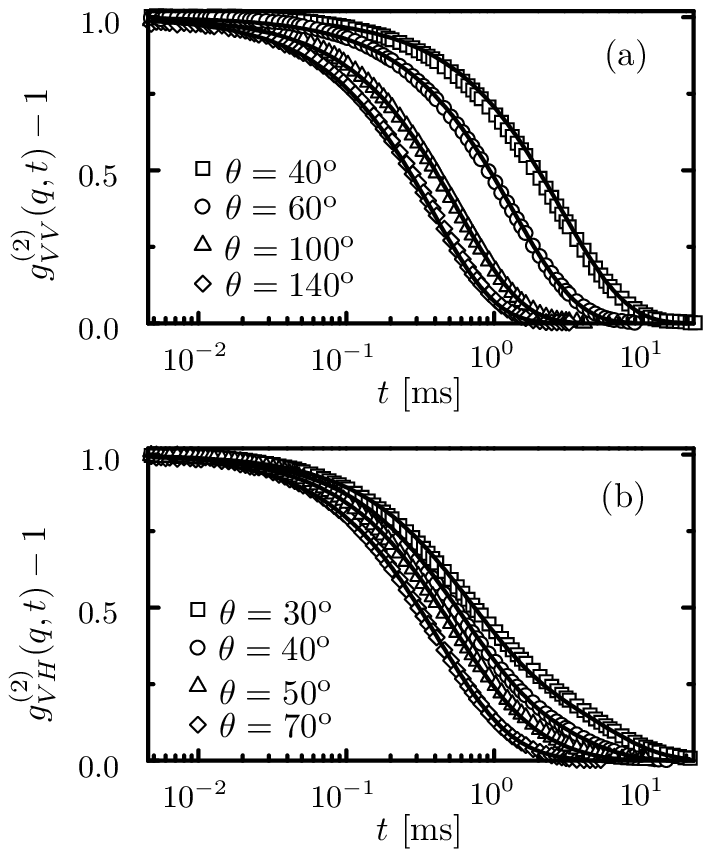}
\vspace*{-0.4cm}
\caption{(a) [(b)] DLS [DDLS] intensity autocorrelation functions $g^{(2)}_{VV}(q,t)$ 
[$g^{(2)}_{VH}(q,t)$] (symbols) of the core-shell particles containing palladium 
nanoparticles at \mbox{$T=25\,^{\rm o}$C} (see Figure \ref{fig1} (b)) together with the 
theoretical results (lines) as obtained from eqs \ref{eq1} - \ref{eq4} with 
\mbox{$a=115$ nm} and \mbox{$\alpha=1.6$}. In (a) and (b) the scattering
angle $\theta$ increases from right to left.}
\label{fig3}
\end{center} 
\end{figure}

Figure \ref{fig3} displays examples of measured and calculated intensity autocorrelation 
functions of the core-shell particles containing palladium nanoparticles at 
\mbox{$T=25\,^{\rm o}$C}. In the calculations the model parameters 
\mbox{$\eta=0.896\times 10^{-3}$ Ns/m$^2$}, \mbox{$a=115$ nm}, and 
\mbox{$\alpha=1.6$} have been used. The radius agrees with the radius of the particles 
as obtained from the CryoTEM micrographs shown in Figures \ref{fig1} (a) and (b). However, 
it turned out that a value of \mbox{$\alpha=1.6$} above unity had to be chosen in order to 
describe the experimental data. The value \mbox{$\alpha=1.6$} found here indicates that the 
core-shell particles exhibit additional dynamics resulting from the shape fluctuations shown 
in Figures \ref{fig1} (a) and (b). The difference between the measured DDLS data and the 
ones calculated from the hard sphere model increase upon decreasing the temperature. This 
is illustrated in Table \ref{tab1}, where the parameter $\alpha$ is presented for three 
temperatures.

%
%
%
%

%
%
\begin{table}[t!]
\begin{tabular}{|c||c|c|c|}\hline
temperature $T$ &\mbox{45\,$^{\rm o}$C} & \mbox{25\,$^{\rm o}$C} 
& \mbox{15\,$^{\rm o}$C}\\\hline\hline
radius $a$ [nm] & 78 & 115 & 128 \\\hline
$\alpha$& 1 & 1.6 & 2.5\\\hline
$D_R(a,\alpha)$ [s$^{-1}$]& 611 & 192 & 168\\\hline
\end{tabular}
\caption{The radius $a$ of the core-shell particles containing palladium
nanoparticles, the parameter $\alpha$, and the diffusion coefficient $D_R(a,\alpha)$ 
as obtained from modelling experimental scattering data in terms of 
eqs \ref{eq1} - \ref{eq4}. Above the transition temperature \mbox{$T=32\,^{\rm o}$C} 
the dynamics of the core-shell particles is the same as those for hard spheres ($\alpha=1$) 
because the shell provides a rather tight envelope of the core (see Figure \ref{fig1} (c)). 
Below the transition temperature the particles exhibit additional dynamics resulting 
from shape fluctuations ($\alpha > 1$ and see Figures \ref{fig1} (a) and (b)).}
\label{tab1}
\end{table}
%
%

\begin{figure}[t!]
\begin{center}
\includegraphics[width=7.5cm,clip]{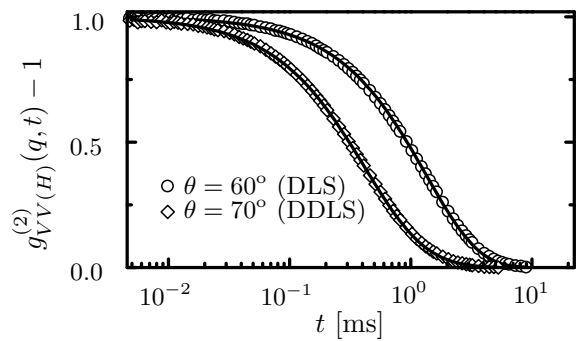}
\caption{Measured DLS (circles) and DDLS (diamonds) intensity autocorrelation functions 
$g^{(2)}_{VV}(q,t)$ and $g^{(2)}_{VH}(q,t)$ of the core-shell particles without embedded 
palladium nanoparticles at \mbox{$T=25\,^{\rm o}$C} (see Figure \ref{fig1} (a))
together with the same theoretical results (lines) used in Fig.~\ref{fig3} for the 
core-shell particles containing palladium nanoparticles at \mbox{$T=25\,^{\rm o}$C}.}
\label{fig4}
\end{center} 
\end{figure}

We have found that embedding nanoparticles in the network of the shell only weakly 
influences the dynamics of the core-shell particles as is apparent from Fig.~\ref{fig4}.
In this figure examples of measured DLS and DDLS intensity autocorrelation functions of the 
core-shell particles without embedded palladium nanoparticles at \mbox{$T=25\,^{\rm o}$C} 
are shown together with the theoretical results (lines) already used in Fig.~\ref{fig3} that 
refer to core-shell particles containing palladium nanoparticles. Both measurements refer 
to the same temperature and the comparison demonstrates that the embedded nanoparticles 
do not disturb the volume transition of the thermosensitive network. No additional 
crosslinking or influence of the nanoparticles on the polymer chains in the network 
is seen in full accord with previous findings.\cite{lu:06,lu:07b}

%
%
%
%

\subsection{Slow and fast mode}

\begin{figure}[t!]
\begin{center}
\includegraphics[width=7.5cm,clip]{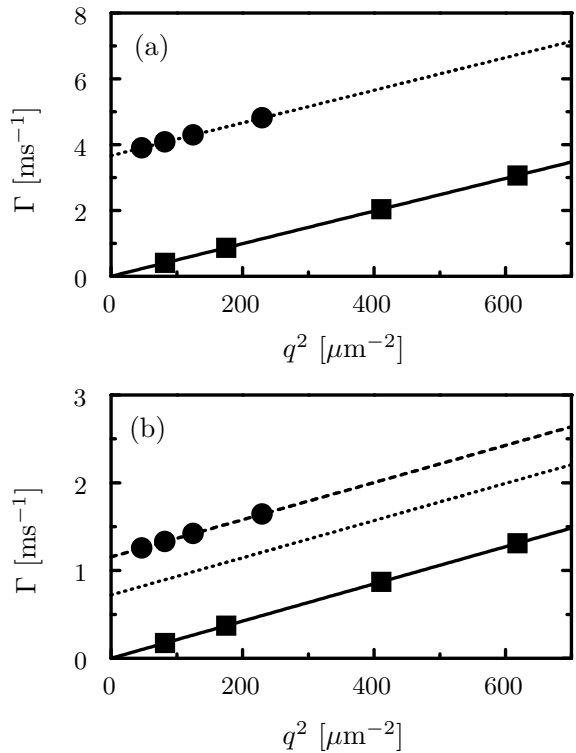}
\vspace*{-0.4cm}
\caption{Decay rates $\Gamma_{slow}(q,a)$ (solid lines), $\Gamma_{fast}(q,a,\alpha=1)$ 
(dotted lines), and $\Gamma_{fast}(q,a,\alpha=1.6)$ (dashed line) in (b) as 
calculated according to eqs \ref{eq10} and \ref{eq11}. The squares and circles 
denote the slow and fast relaxation rate, respectively, used to describe the 
intensity autocorrelation functions of the core-shell particles containing palladium 
nanoparticles (see Figures \ref{fig2} and \ref{fig3}) at \mbox{$T=45\,^{\rm o}$C} in 
(a) and \mbox{$T=25\,^{\rm o}$C} in (b). The absolute value of the scattering vector 
is given by $q=26.45\, \sin(\theta/2)\,\mu$m$^{-1}$ where $\theta$ is the scattering 
angle. Hence $q^2=618\,\mu$m$^{-2}$ corresponds to $\theta=140\,^{\rm o}$.}
\label{fig5}
\end{center} 
\end{figure}

The autocorrelation function $\sqrt{g^{(2)}_{VH}(q,t)-1}$ in eq \ref{eq2} is a
sum of two discrete exponentially decaying functions, where the slow relaxation 
mode characterizes translational diffusion while the faster relaxation mode is 
related to rotational motion and shape fluctuations. Hence, one may describe the 
experimental data in terms of a slow ($\Gamma_{slow}$) and a fast relaxation rate
($\Gamma_{fast}$) according to
\begin{eqnarray} \label{eq10}
\Gamma_{slow}(q,a)&=& q^2 D_T(a)\,,
\\\Gamma_{fast}(q,a,\alpha)&=& q^2 D_T(a)+6 D_R(a,\alpha)\,. \label{eq11}
\end{eqnarray}
The slow mode is also characteristic for the single exponential decay of the 
autocorrelation function $\sqrt{g^{(2)}_{VV}(q,t)-1}$ in eq \ref{eq1}. 
Figure \ref{fig5} shows the calculated relaxation rates as a function of $q^2$.
The solid and dotted lines denote $\Gamma_{slow}(q,a)$ and $\Gamma_{fast}(q,a,\alpha=1)$, 
respectively, where \mbox{$a=78$ nm} at \mbox{$T=45\,^{\rm o}$C} in (a) and  
\mbox{$a=115$ nm} at \mbox{$T=25\,^{\rm o}$C} in (b). The hard sphere model,
i.e., $\alpha=1$, is indeed appropriate for the microgel particles at 
\mbox{$T=45\,^{\rm o}$C} as discussed above and apparent from a comparison 
of the calculated results (solid and dotted lines) with the experimental 
data (symbols) in Figure \ref{fig5} (a). The pronounced differences between the 
calculated decay rates $\Gamma_{fast}(q,a,\alpha=1)$ (dotted line) and the 
experimental data (circles) in Figure \ref{fig5} (b) are due to the shape 
fluctuations at \mbox{$T=25\,^{\rm o}$C}. Therefore, the value $\alpha=1.6$
(see Table \ref{tab1}) has been used in computing the dashed line 
displaying $\Gamma_{fast}(q,a,\alpha=1.6)$.

%
%
%
%

We emphasize that neither modeling the core-shell particles as hard nonspherical
particles such as ellipsoids or dimers nor taking into account the small size polydispersity of 
the particles or intermolecular interactions lead to an agreement with the experimental 
data. For example, the corresponding diffusion coefficients of a hard dimer consisting 
of two identical hard spheres of radius $a$  are given by $D_T^{(dim)}(a)=0.718\, D_T(a)$ 
and $D_R^{(dim)}(a)=0.265\, D_R(a)$, respectively. \cite{carr:99} Hence the ratio of 
the rotational diffusion coefficient to the translational diffusion coefficient 
$D_R/D_T$ is smaller for hard dimers than for hard spheres, while the opposite behavior 
has been found for the core-shell particles below the transition temperature. 

Size polydispersity 
of the particles leads to a considerably slower decay of the intensity autocorrelation functions
\cite{harn:99} in comparison with both the corresponding autocorrelation functions of a 
monodisperse system and the experimental data. Moreover, $D_T$ and $D_R$ decrease with 
increasing volume fraction of hard spheres. \cite{degi:91} Hence, this comparison demonstrates 
clearly that the dynamics of the core-shell particles below the transition temperature cannot
be explained in terms of a hard particle model. 

Furthermore, the experimental results cannot be explained by using slipping boundary conditions 
(see, e.g., ref\cite{hu:74}) instead of the conventional sticking boundary conditions which 
lead to eqs \ref{eq3} an \ref{eq4}. The translational diffusion coefficient of a sphere with 
slipping boundary conditions is increased by the factor $3/2$ as compared to $D_T(a)$ in 
eq \ref{eq3}, while the rotational motion of such a sphere does not displace any fluid, 
implying that $D_R(a) \to \infty$. Both results do not agree with the experimental data.

\subsection{Coupling of the shape fluctuations with rotational motion }

\begin{figure}[t!]
\begin{center}
\includegraphics[width=7.5cm,clip]{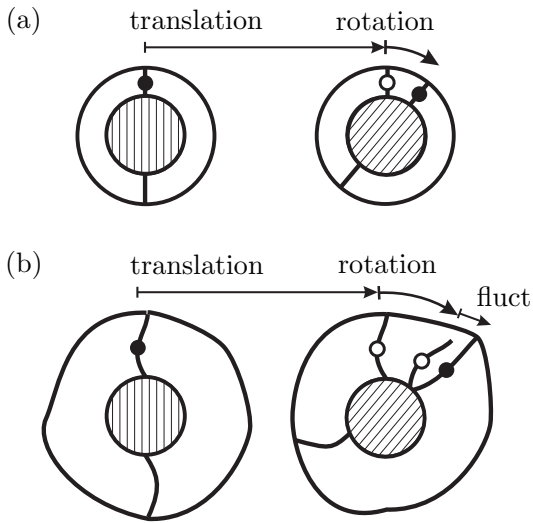}
\caption{Illustration of the motion of a scattering unit (solid circle) on a PNIPA 
chain within the shell of the core-shell particles, where the hatched 
region marks the core of the particles. The scattering unit exhibits translational 
motion and rotational motion around the center of mass of the particle in the case 
of a rather tight shell in (a), while there is an additional fluctuation dynamics 
(fluct) due to internal degrees of freedom of the PNIPA chain if the shell is rather 
soft in (b).}
\label{fig6}
\end{center} 
\end{figure}

In the following we present a model that takes into account the influence of shape fluctuations 
on the rotational motion of the particles. As is illustrated in Figure \ref{fig6},
the  motion of a scattering unit (solid circle) on a PNIPA chain within the shell of the 
particles can be decomposed into various types of modes. If the shell provides a rather 
tight envelop of the core, the scattering unit will exhibit translational motion 
and rotational motion around the center of mass of the particle (Figure \ref{fig6} (a)), that is, 
the particles will behave as hard spheres without internal degrees of freedom. So above the 
LCST the DDLS signal will be due to the inhomogeneity of the frozen polymer network. However, 
there is an additional fluctuation dynamic (fluct; see Figure \ref{fig6}) due to internal 
degrees of freedom of the PNIPA chain if the shell is rather soft (Figure \ref{fig6} (b)). 
%
%
%
%

In the following we shall discuss the coupling between 
the rotational dynamics and the internal modes of the particles that come into play below 
the transition temperature. The PNIPA chains are linear chain molecules which are described by 
a chain model for macromolecules of variable stiffness \cite{harn:95,harn:96} that has been 
used earlier to discuss the dynamics of polymers under the influence of various forces 
(see, e.g., refs \cite{harn:99a,harn:99b,harn:99c,wink:06} and references therein). 
It has been shown that the dynamics of individual PNIPA chains in dilute solution can be 
interpreted in terms of a chain model of this type. \cite{wu:96} We consider a continuous,
differentiable space curve ${\bf r}(s,t)$ inscribed into a sphere of radius $a$, where 
$s \in[-a,a]$ is a coordinate along the macromolecule and ${\bf r}(0,t)$ is the position 
vector of the center of the sphere (see Figure \ref{fig6}). The potential energy functional 
of the model reads \cite{harn:95,harn:96} 
\begin{eqnarray} \label{eq5}
\lefteqn{U_{pot}[{\bf r}(s,t)]=}\nonumber
\\&&\int\limits_{-a}^ads\,
\left[\nu p(s)\left(\frac{\partial {\bf r}(s,t)}{\partial s}\right)^2
+\frac{\epsilon}{p(s)} \left(\frac{\partial {\bf r}^2(s,t)}{\partial s^2}\right)^2
\right]\nonumber
\\&&+\nu_0 \left[\left(\frac{\partial {\bf r}(-a,t)}{\partial s}\right)^2
+\left(\frac{\partial {\bf r}(a,t)}{\partial s}\right)^2\right]\,,
\end{eqnarray}
where $1/p(s)$ is a local correlation length characterizing the stiffness of the space 
curve and $\nu_0$, $\nu$, $\epsilon$ are Lagrange multipliers.
In the limit $p(s) \to 0$, the space curve describes a rigid vector inscribed into 
a hard sphere which exhibits translational and pure rotational Brownian motion. Internal 
fluctuation of the PNIPA chains are taken into account in terms of $p(s) \neq 0$ for values 
of $s$ inside the shell. The term with the first derivative in ${\bf r}(s,t)$ captures the 
chain flexibility, i.e., it takes chain entropy into account. The term with the second 
derivative accounts for bending stiffness and the last two terms are due to the broken 
symmetry at the chain ends \cite{harn:95}. In order to gain analytical insight we consider 
the case that the flexibility parameter $p(s)$ does not depend on $s$, that is, $p(s)= p$. 
Applying Hamilton's principle we find the Langevin equation of motion along with the boundary conditions for free ends,
\begin{eqnarray} 
3\pi\eta\frac{\partial}{\partial t}{\bf r}(s,t)-
2\nu p \frac{\partial^2}{\partial s^2}{\bf r}(s,t)+
\frac{\epsilon}{p} \frac{\partial^4}{\partial s^4}{\bf r}(s,t)
&=&{\bf f}(s,t)\,,\nonumber
\\\label{eq13}
\\\left[2\nu p \frac{\partial}{\partial s}{\bf r}(s,t)-
\frac{\epsilon}{p} \frac{\partial^3}{\partial s^3}{\bf r}(s,t)\right]_{\pm a}
&=&0\,,
\\\left[2\nu_0 \frac{\partial}{\partial s}{\bf r}(s,t)+
\frac{\epsilon}{p} \frac{\partial^2}{\partial s^2}{\bf r}(s,t)\right]_{a}
&=&0\,,
\\\left[2\nu_0 \frac{\partial}{\partial s}{\bf r}(s,t)-
\frac{\epsilon}{p} \frac{\partial^2}{\partial s^2}{\bf r}(s,t)\right]_{-a}
&=&0\,,
\end{eqnarray}
where ${\bf f}(s,t)$ is the stochastic force. The first term in eq \ref{eq13} 
represents the frictional force. Equation \ref{eq13} is a fourth-order, linear 
partial differential equation which can be solved by means of a normal mode analysis.
The eigenvalue problem including the boundary conditions is hermitian. Therefore,
the eigenfunctions $\psi_l(s)$ are orthogonal and form a complete set. An expansion
of the position vector and of the stochastic force in terms of the eigenfunctions
and the time-dependent amplitudes $\mbox{\boldmath$\chi$}_l(t)$, ${\bf f}_l(t)$
according to ${\bf r}(s,t)=\sum_{l=0}^\infty \psi_l(s)\mbox{\boldmath$\chi$}_l(t)$
and ${\bf f}(s,t)=\sum\limits_{l=0}^\infty \psi_l(s){\bf f}_l(t)$ yields
\begin{eqnarray} 
{\bf r}(s,t)&=&\sum\limits_{l=1}^\infty \frac{\psi_l(s)}{3\pi\eta}
\int\limits_{-\infty}^t dt'\,{\bf f}_l(t') e^{-(t-t')/\tau_l}\nonumber
\\&&+
\frac{\psi_0}{3\pi\eta}\int\limits_{-\infty}^t dt'\,{\bf f}_0(t')\,.
\end{eqnarray}
Here $3\pi\eta/\tau_l$ is the eigenvalue corresponding to the eigenfunction 
$\psi_l(s)$. The eigenfunction $\psi_0=1/\sqrt{2a}$ belongs to the eigenvalue 
$3\pi\eta/\tau_0=0$ and corresponds to the translational motion 
of the center of mass
\begin{eqnarray} 
{\bf r}_{cm}(t)&\equiv&\frac{1}{2a}\int\limits_{-a}^a ds\,{\bf r}(s,t)=
\frac{\psi_0}{3\pi\eta}\int\limits_{-\infty}^t dt'\,{\bf f}_0(t')\,,
\end{eqnarray}
because $\int_{-a}^ads\, \psi_l(s)=0\,\,\,, \forall\,\, l\neq 0$. Assuming Gaussian
distributed random forces ${\bf f}(s,t)$ characterized by the thermal average
$\langle f_n(s,t) \rangle = 0$ and 
\begin{eqnarray} 
\left\langle  f_n(s,t) f_m(s',t') \right\rangle&=& 6 \pi \eta k_BT 
\delta_{nm}\delta(s-s')\delta(t-t')\,,\nonumber
\\&&
\,\,\,\, n,m\in\{x,y,z\}
\end{eqnarray}
the translational diffusion coefficient $D_T(a)$ is of the form of eq \ref{eq3}:
\begin{eqnarray} 
D_T(a)&\equiv& \lim_{t\to \infty} \frac{1}{6t}\left\langle \left(
{\bf r}_{cm}(t)-{\bf r}_{cm}(0)\right)^2\right\rangle\nonumber
\\&=&
\frac{k_BT}{\eta}\frac{1}{6\pi a}\,.
\end{eqnarray}
Hence $D_T(a)$ is independent of the local correlation length $1/p$ which is valid 
in the so-called free-draining limit for dense polymer systems 
such as microgels. \cite{harn:95,harn:99a,harn:99b}
On the other hand, intramolecular hydrodynamic 
interactions lead to a dependence of $D_T(a)$ on $1/p$ in the case of  dilute 
\cite{harn:99,harn:96,harn:99c} or semi-dilute \cite{harn:01,boli:07} polymer solutions.

\begin{figure}[t!]
\begin{center}
\includegraphics[width=7.5cm,clip]{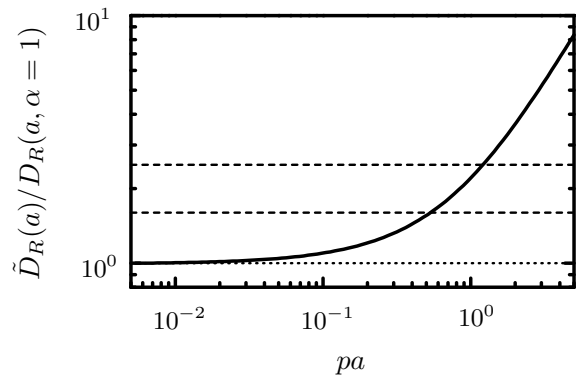}
\caption{The reduced rotational-fluctuation diffusion coefficient \mbox{${\tilde D}_R(a)/
D_R(a,\alpha=1)$} according to eqs \ref{eq7} and \ref{eq4} as a function of  $pa$ 
characterizing the stiffness of the model described in the main text. In the stiff 
limit $pa\to 0$ the rotational-fluctuation diffusion coefficient ${\tilde D}_R(a)$ reduces 
to the rotational diffusion coefficient \mbox{$D_R(a,\alpha=1)$} of hard spheres. The 
deviations of \mbox{${\tilde D}_R(a)/D_R(a,\alpha=1)$} from the value $1$ increase with  
decreasing stiffness, i.e., increasing values of $pa$, due to additional internal dynamics.
The dashed lines mark the values $\alpha=1.6$ and $\alpha=2.5$ as obtained from 
the experimental data at \mbox{$T=25\,^{\rm o}$C} and \mbox{$T=15\,^{\rm o}$C}
(see Table \ref{tab1}).}
\label{fig7}
\end{center} 
\end{figure}

The first internal mode ($l=1$) exhibits the largest relaxation time and the 
rotational-fluctuation diffusion coefficient ${\tilde D}_R(a)=1/(3\tau_1)$ 
can be derived as 
\begin{eqnarray} \label{eq7}
{\tilde D}_R(a)&=&\frac{k_BT}{\eta}\frac{\alpha_1^4+4\alpha_1^2 p^2}{48\pi p}
\\&\approx&
\left\{\begin{array}{c@{\quad,\quad}l}
\frac{\displaystyle k_BT}{\eta}\frac{\displaystyle 1}{\displaystyle 8\pi a^3}& pa\lesssim 0.02
\\\label{eq8}
\frac{\displaystyle k_BT}{\eta}\frac{\displaystyle \pi p}{\displaystyle 48 a^2}& pa\gtrsim 2
\end{array}\right.\,,
\end{eqnarray}
where $\alpha_1$ follows from the transcendental equation 
\begin{eqnarray} 
&&\alpha_1^3\sin(\alpha_1 a)\,\cosh(\beta_1 a)
-\beta_1^3\cos(\alpha_1 a)\,\sinh(\beta_1 a)\nonumber
\\&&-2p(\alpha_1^2+\beta_1^2)
\cos(\alpha_1 a)\,\cosh(\beta_1 a)=0
\end{eqnarray}
together with \mbox{$\beta_1^2-\alpha_1^2=4p^2$} and $\nu_0=3/(16 k_BT)$, $\nu=3/(8 k_BT)$, 
$\epsilon_0=3/(16 k_BT)$. The normalized rotational-fluctuation diffusion 
coefficient ${\tilde D}_R(a)/D_R(a,\alpha=1)$ is plotted in Figure \ref{fig7} as function 
of $pa$. In the stiff limit $pa\to 0$ the diffusion coefficient ${\tilde D}_R(a)$ agrees 
exactly with the rotational diffusion coefficient \mbox{$D_R(a,\alpha=1)$} of hard spheres. 
With decreasing stiffness (increasing values of $pa$) the ratio 
${\tilde D}_R(a)/D_R(a,\alpha=1)$ increases similar to the experimental findings presented in 
Table \ref{tab1}. Hence shape fluctuations do indeed lead to a considerable increase 
of the diffusion coefficient ${\tilde D}_R(a)$. \\
Of course, the results shown 
in Figure \ref{fig7} can only be considered to be of qualitative significance for the 
core-shell particles under consideration. However, these results clearly point to the 
importance of the coupling of rotational and internal modes in the case of soft materials. 
We emphasize that the core-shell particles maintain on average a spherical shape because
$g^{(2)}_{VV}(q,t)-1$ can be described by a single exponential function according to 
eq \ref{eq1}. Any permanent deviation from a spherical shape would lead to an additional 
term in eq \ref{eq1} due to rotational motion.

%
%
%
%
\vspace*{1cm}

Finally, it is worthwhile to compare the results of the present investigation with earlier 
studies of properties of microgels. A key feature of the static scattering intensity of 
both uniform microgel particles and core-shell microgel particles below the transition 
temperature is a strong scattering signal that is due to collective fluctuations of the 
polymer gel. \cite{ding:98,seel:01,krat:01} It has been demonstrated that the contribution 
of this internal dynamics to the static scattering intensity vanishes for temperatures 
above the transition temperature \mbox{T=32\,$^{\rm o}$C} (see Figure 4 in ref~\cite{krat:01}). 
The corresponding collective dynamic fluctuations  of the PNIPA network
below the LCST have been measured using DLS \cite{wu:96,yang:04} and neutron spin-echo 
spectroscopy. \cite{hell:02} Despite the different $q$ range of the two methods, the found 
collective diffusion coefficient of the network has a similar order of magnitude. This 
most likely indicates that the network motion is always observable independent of the 
characteristic length scale of the experiment. Hence, the deviations from the hard sphere 
model observed in the present study by means of DDLS can be considered as a manifestation 
of these fluctuations. Very recently it has been demonstrated experimentally that the softness 
of microgel particles has also a pronounced influence on the dynamics in concentrated 
microgel suspensions. \cite{ecke:08}

\section{Conclusion}
In conclusion, our findings elucidate an important and interesting interplay between shape 
fluctuations and rotational motion of deformable objects which profoundly affects their 
dynamics. The control over the degree of deformations 
offered by varying the temperature should make the core-shell microgels useful 
for fundamental studies in statistical physics. We anticipate that the results 
obtained for the present system is of general importance for a better understanding 
for more complicated systems related to biophysics as e.g. vesicles.

\section{Acknowledgment}
Financial support by the Deutsche Forschungsgemeinschaft, SFB 481, Bayreuth, and by the 
Schwerpunktprogramm Hydrogele is gratefully acknowledged.

\end{document}